\documentclass[prd,notitlepage,nofootinbib,preprintnumbers,amssymb,superscriptaddress]{revtex4-1}
\usepackage{amsfonts,amssymb,amsmath,graphicx,color,bm,ulem}
\usepackage{latexsym,subfigure,ascmac}
\usepackage[Symbolsmallscale]{upgreek} % Euler(default), Symbol, Symbolsmallscale
\usepackage[euler]{textgreek} % cbgreek(default), euler, artemisia
\definecolor{ultramarine}{rgb}{0.07, 0.04, 0.56}
\definecolor{cadmiumgreen}{rgb}{0.0, 0.42, 0.24}
\definecolor{indigo(dye)}{rgb}{0.0, 0.25, 0.42}
\usepackage[linktocpage=true]{hyperref}
\hypersetup{
colorlinks=true,
citecolor=ultramarine,
linkcolor=cadmiumgreen,
urlcolor=indigo(dye),
}

\newcommand{\f}[2]{\frac{#1}{#2}}  
\newcommand{\mk}[1]{\left( #1 \right)}

\newcommand{\be}{\begin{equation}}  
\newcommand{\ee}{\end{equation}}
\newcommand{\lamas}{\alpha}
\newcommand{\KJ}{{\tilde K}}
\newcommand{\Hal}{\upalpha}
\newcommand{\Hbe}{\upbeta}
\newcommand{\Hga}{\upgamma}
\newcommand{\Hde}{\updelta}
\newcommand{\Hep}{\upvarepsilon}
\newcommand{\mR}{\mathcal{R}}
\newcommand{\mT}{\mathcal{T}}
\newcommand{\mO}{\mathcal{O}}
\renewcommand{\a}{{\rm a}}
\renewcommand{\c}{{\rm c}}
\newcommand{\q}{{\rm q}}

\begin{document}%%%%%%%%%%%%%%%%%%%%%%%%%%%%%%%%%%%%%%%%%  

\title{Evaporation of Echoing Black Holes}

\author{Naritaka Oshita}
\affiliation{RIKEN iTHEMS, Wako, Saitama, 351-0198, Japan}

\author{Hayato Motohashi} 
\affiliation{Division of Liberal Arts, Kogakuin University, 2665-1 Nakano-machi, Hachioji, Tokyo, 192-0015, Japan}

\author{Sousuke Noda}
\affiliation{National Institute of Technology, Miyakonojo College, Miyakonojo 885-8567, Japan}

\preprint{RIKEN-iTHEMS-Report-22}

\begin{abstract}%%%%%%%%%%%%%%%%%%%%%%%%%%%%%%%%%%%%%%%%%
We compute the graybody factor and evaporation rate of a rotating black hole in the presence of a hypothetical reflective surface slightly outside the outer horizon radius, assuming that it spontaneously emits thermal radiation 
due to quantum-gravitational effects such as firewalls or stretched horizons. 
As a result of a resonance caused by a cavity between the reflective surface and angular momentum barrier, the graybody factor is subject to a modulation in the frequency space.
By taking into account this effect for multiangular modes of neutrinos, photons, and gravitons,
we numerically compute the time development of the mass and angular momentum of the black hole, and show that 
the excited reflective surface shortens the lifetime of quantum black holes.
\end{abstract}%%%%%%%%%%%%%%%%%%%%%%%%%%%%%%%%%%%%%%%%%

\maketitle  

%%%%%%%%%%%%%%%%%%%%%%%%%%%%%%%%%%%%%%%%%  

\section{Introduction}%%%%%%%%%%%%%%%%%%%%%%%%%%%%%%%%%%%%%%%%%
Black holes are among the most mysterious objects in the Universe, and their classical and quantum-mechanical aspects have been actively investigated for several decades. Although a classical black hole is a perfect absorber and emits nothing out of the near-horizon region, an actual black hole may induce vacuum polarization around the horizon, which causes the emission of thermal Hawking particles and leads to the evaporation of the black hole~\cite{Hawking:1974rv,Hawking:1975vcx}. Therefore, the lifetime of a black hole is determined by the emission rate of Hawking radiation.

The spectrum or its emission rate of Hawking radiation may be sensitive to the near-horizon structure of a black hole. Despite intense efforts to theoretically understand the evaporation process of black holes, a high-precision probe of the near-horizon region of a black hole has not been achieved yet. Nevertheless, there is some tentative evidence implying the possibility that the near-horizon structure differs from the classical one due to quantum-gravitational effects. One of the most topical issues is the tentative evidence of gravitational wave (GW) echoes that may appear at late times in the GW signal from the binary black hole merger events. Searching for the GW echoes was pioneered by Ref.~\cite{Abedi:2016hgu} and many other searches for echoes have been performed; however, depending on the methodologies used, they found positive, mixed, or negative evidence~\cite{Ashton:2016xff,Conklin:2017lwb,Westerweck:2017hus,Abedi:2018pst,Abedi:2018npz,Conklin:2019fcs,Uchikata:2019frs,Abedi:2021tti}. If the GW echoes are indeed emitted after ringdown from a remnant black hole, it might be a footprint of quantum-gravitational effects at the near-horizon region~\cite{Cardoso:2016rao,Cardoso:2016oxy,Holdom:2016nek,Oshita:2018fqu,Cardoso:2019apo,Oshita:2019sat,Dey:2020lhq,Maggio:2020jml}.

Recently, many theoretical models that may cause the emission of the GW echoes have been proposed and investigated from a theoretical point of view~\cite{Cardoso:2016rao,Cardoso:2016oxy,Holdom:2016nek,Oshita:2018fqu,Cardoso:2019apo,Oshita:2019sat,Dey:2020lhq,Maggio:2020jml,Cardoso:2014sna,Nakano:2017fvh,Maggio:2017ivp,Cardoso:2017cqb,Wang:2018gin,Testa:2018bzd,Cardoso:2019rvt,Wang:2019rcf,Oshita:2020dox,Oshita:2020abc,Sago:2020avw,LongoMicchi:2020cwm,Sago:2022bbj}, and the numerical relativity is also employed to compute the waveform of the GW echoes~\cite{Ma:2022xmp}. In most models, a reflective surface at the near-horizon region is involved. The GW echoes are emitted due to the resonance between the reflective surface and the angular momentum barrier. The surface is usually characterized by two quantities: its reflectivity and position of the boundary. As a simplified model of the reflection, the reflectivity is assumed to be constant with respect to the frequency of incoming waves. As another model, the Boltzmann reflectivity, for which the reflectivity is given by the Boltzmann factor with the Hawking temperature, has been proposed~\cite{Oshita:2018fqu,Oshita:2019sat,Wang:2019rcf} and its significance on the GW signal has been investigated~\cite{Abedi:2020ujo,Oshita:2020dox,Oshita:2020abc,Sago:2020avw,LongoMicchi:2020cwm,Ma:2022xmp}.

\begin{figure}[t]
  \begin{center}
    \includegraphics[keepaspectratio=true,height=90mm]{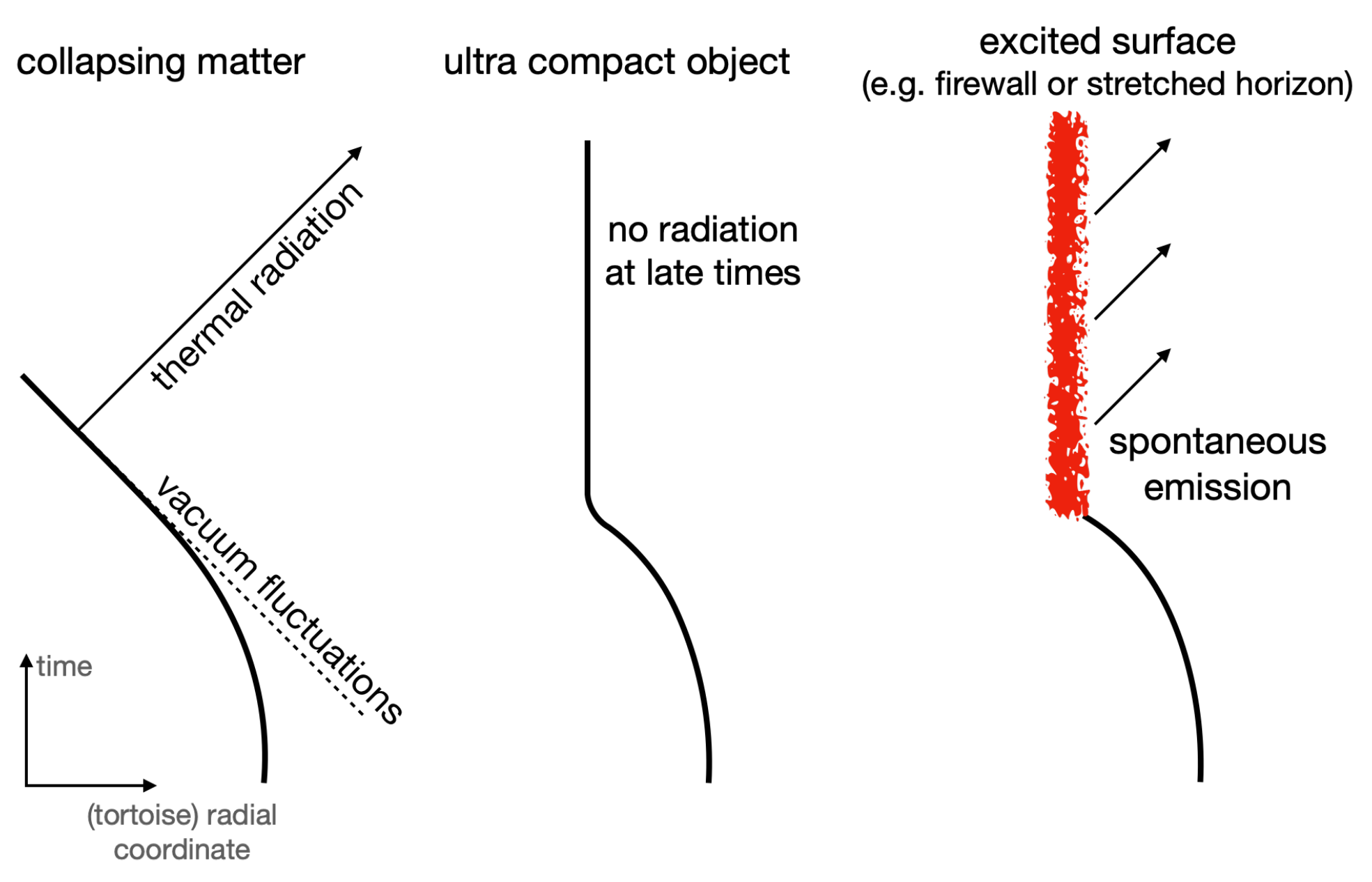}
  \end{center}
\caption{Three scenarios of the emission process out of a black hole. Vacuum fluctuations interacting with collapsing matter near the horizon lead to thermal radiation (left), which is the original proposal by Hawking \cite{Hawking:1974rv,Hawking:1975vcx}. For an ultracompact object (middle), there is no radiation at late times as the surface is static and there is no squeezing of vacuum fluctuations \cite{Harada:2018zfg,Kokubu:2019jdx}. We discuss a situation where a black hole spontaneously emits thermal radiation while its surface has a nonzero reflectivity as another possibility (right).
}
\label{three_scenarios}
\end{figure}

In addition to the reflection of the incoming waves, the reflective surface itself may or may not emit thermal radiation {\it spontaneously} as depicted in Fig.~\ref{three_scenarios}.
In Refs.~\cite{Harada:2018zfg,Kokubu:2019jdx}, it is assumed that gravitational collapse leads to the formation of an ultracompact object with a reflective surface, and the particle creation around the object is studied there. In this case, the collapse of the cold boundary eventually terminates and the thermal emission (particle creation) is quenched at late times\footnote{Recently, particle emission induced by the formation of a gravastar was investigated in Ref.~\cite{Nakao:2022ygj}. In that paper, it was reported that the emission of thermal radiation, whose temperature approaches to the Gibbons-Hawking temperature of the de Sitter interior, could last much longer than the free-fall time of the system. Although the radiation would terminate at very late times even in that model \cite{Nakao:2022ygj}, the transient thermal radiation may arise due to the formation of a gravastar.}.
On the other hand, the near-horizon structure of a quantum black hole could involve an energetically or thermally excited surface due to quantum-gravitational effects, e.g.,\ a firewall~\cite{Almheiri:2012rt}, membrane~\cite{Thorne:1986iy}, or stretched horizon~\cite{Susskind:1993if}. It may have its own degrees of freedom, possibly leading to the Bekenstein-Hawking entropy~\cite{Susskind:1993if,Almheiri:2012rt}, and may spontaneously emit thermal particles, unlike a cold boundary.
%%%%%%%%%%%%%%%%%%%%%%%%%
%%%%%%%%%%%%%%%%%%%%%%%%%

In this paper, we consider an evaporation model for a rotating black hole where the (partially) reflective surface is in excited states and spontaneously emits thermal radiation\footnote{We do not expect that the surface has perfect reflectivity for all frequencies as the surface would not be thermalized in that case. Nevertheless, we will consider a nearly perfect reflectivity as an example later to estimate a lower bound on the lifetime of an echoing black hole.}. It amounts to a change of the boundary condition and hence affects the graybody factor. We show that the spontaneous emission of thermal radiation and the resonance between the reflective surface and angular momentum barrier shorten the lifetime of the black hole. This is an interesting possibility that may lead to a totally different evaporation process from the standard one in the context of quantum gravitational phenomenology.

We first compute the graybody factor for a rotating black hole by taking into account a resonance caused by the cavity between the reflective surface and angular momentum barrier. We then compute the mass-loss and spin-loss rates with respect to the reflectivity and the position of the reflective surface. Finally, we simulate the evaporation of the black hole by taking into account multiple harmonic modes for several species of emitted particles such as neutrinos, photons, and gravitons\footnote{Note that as an approximation, neutrinos are regarded as massless spin-half particles in our computation. This is valid when the Hawking temperature is higher than the neutrino mass (for a relevant discussion, see, e.g., Ref.~\cite{Page:1976ki}). For the sake of brevity, we call the massless spin-half particles ``neutrinos."}. Our methodology is based on the previous work that simulates the evaporation of a spinning black hole~\cite{Page:1976ki}. The computation performed in this paper employs the general Heun's function, which represents the exact solution for the perturbations of a test field in the Kerr-de Sitter spacetime~\cite{Suzuki:1999nn}. This analysis has the advantage of avoid the direct numerical integration of the Teukolsky~\cite{Teukolsky:1973ha}, Sasaki-Nakamura~\cite{Sasaki:1981sx}, or Chandrasekhar-Detweiler equation~\cite{Chandrasekhar:1976zz} by virtue of the fact that the Teukolsky equations can be transformed to the Heun's differential equations \cite{Suzuki:1999nn}. It has been employed in the computation of quasinormal modes~\cite{Hatsuda:2020sbn,Oshita:2021iyn}, wave scattering problem~\cite{Motohashi:2021zyv}, and Hawking radiation~\cite{Gregory:2021ozs,Nambu:2021eqe} without the reflective surface. Also, the regular singularities of the general Heun's function enable us to analytically compute the graybody factors~\cite{CarneirodaCunha:2015qln,Novaes:2018fry} at the cost of having a small cosmological constant. In Appendix \ref{sec:app}, we check that the effect of the small cosmological constant we use in our simulation is negligible. We use natural units, where $c=\hbar=G=k_{\rm B}= 1$, throughout the paper.

\section{Exact solution}%%%%%%%%%%%%%%%%%%%%%%%%%%%%%%%%%%%%%%%%%
\label{sec:sol}
In this section, we briefly review the fact that the Teukolsky equation for the Kerr-de Sitter solution can be transformed to the form of the Heun's differential equations. The Heun's differential equation, obtained from the radial Teukolsky equation, has analytic solutions near the regular singularities located at the black hole and cosmological horizons. This is the case even for the angular equation for which the regular singularities correspond to the poles. Therefore, we can consider a scattering problem in the Kerr-de Sitter background while keeping the spheroidal harmonics regular by employing the analytic solutions of the Heun's differential equations.

\subsection{Teukolsky equation}%%%%%%%%%%%%%%%%%%%%%

We consider the Kerr-de Sitter spacetime, whose metric in Boyer-Lindquist coordinates is given by
\be ds^2 = - \f{\Delta}{(1+\lamas)^2\rho^2} (dt-a\sin^2\theta d\varphi)^2
+ \rho^2 \mk{ \f{dr^2}{\Delta} + \f{d\theta^2}{1+\lamas \cos^2\theta} } 
+ \f{ (1+\lamas \cos^2\theta)\sin^2\theta }{(1+\lamas)^2\rho^2} [adt-(r^2+a^2) d\varphi]^2
, \ee 
where  
\begin{align} 
\lamas &= \f{\Lambda a^2}{3}, \qquad \rho^2=r^2+a^2 \cos^2\theta, \\
\Delta(r) &=(r^2+a^2)\mk{1-\f{\Lambda}{3}r^2}-2Mr \notag\\
&= -\f{\Lambda}{3}(r-r_-)(r-r_+)(r-r'_+)(r-r'_-).
\end{align} 
Here, $\Lambda$ is a cosmological constant, 
$M$ is a mass parameter, and  
$a$ is a spin parameter.
Throughout the present paper, we assume $\Lambda> 0$, and focus on the case where $\Delta(r)=0$ has four distinct real roots, which are denoted by $r_{\pm},r'_{\pm}$ with the ordering $r'_-<0\leq r_-<r_+<r'_+$.
This assumption is necessary for the formulation in terms of the general Heun function $Hl$.
Here, $r_-$ is the Cauchy horizon, $r_+$ is the event horizon, $r'_+$ is the cosmological horizon, and $r'_-=-(r_-+r_++r'_+)$ holds.
For later convenience, we introduce the tortoise coordinate $r^*$ defined by 
\be \label{drstardef} dr^*=\f{(1+\lamas)(r^2+a^2)}{\Delta(r)}dr, \ee
or 
\be \label{rstar} r^*= \f{\ln|r-r_+|}{2\kappa(r_+)} + \f{\ln|r-r_+'|}{2\kappa(r_+')} + \f{\ln|r-r_-'|}{2\kappa(r_-')} + \f{\ln|r-r_-|}{2\kappa(r_-)} , \ee
where 
\be \kappa(r)=\f{\Delta'(r)}{2(1+\lamas)(r^2+a^2)}. \ee

We focus on the evolution of spin-$s$ massless test fields on the Kerr-de Sitter background.
Decomposing the master variable as
\be \psi_s(t,r,\theta,\varphi) = \int\f{d\omega}{2\pi} e^{-i\omega t}e^{im\varphi} R_s(r) S_s(\theta), \ee
the evolution equation is separated out and takes the unified form for the spin $0, \f{1}{2}, 1, \f{3}{2}, 2$ given by~\cite{Suzuki:1998vy}
\begin{align} 
\label{Teu-ang}
&\Biggl[ \f{d}{d\mu}(1+\lamas \mu^2)(1-\mu^2)\f{d}{d\mu} +\lambda -s(1-\lamas) -2\lamas \mu^2 \notag\\
&~~+ \f{4 s \mu (1 + \lamas) [m \lamas - c (1 + \lamas)]}{1 + \lamas \mu^2}
-\f{(1 + \lamas)^2 [m + s \mu - (1 - \mu^2) c]^2}{(1 + \lamas \mu^2) (1 - \mu^2)} \Biggr] S_s(\theta) = 0, \\
\label{Teu-rad}
&\Biggl[ \Delta^{-s}\f{d}{dr}\Delta^{s+1}\f{d}{dr} 
+ \f{\KJ^2-is\KJ\Delta'}{\Delta} +2 i s \KJ' 
- \f{2}{3}\Lambda r^2(s+1)(2s+1)
+2s(1-\lamas)-\lambda \Biggr] R_s(r) = 0 ,
\end{align}
which are angular and radial Teukolsky equations, respectively. Here, $\lambda$ is a separation constant
and 
\be 
\mu=\cos \theta, \qquad 
c=a\omega, \qquad 
\KJ(r)=(1+\lamas)[\omega(r^2+a^2)-am]. 
\ee

\subsection{General Heun function}%%%%%%%%%%%%%%%%%%%%%

The angular and radial Teukolsky equations can be recast into the general Heun equation~\cite{Suzuki:1998vy}, which takes the form of
\be \label{Heun} \f{d^2f}{d\zeta^2} + \mk{ \f{\Hga}{\zeta} + \f{\Hde}{\zeta-1} + \f{\Hep}{\zeta-\a} } \f{df}{d\zeta} + \f{\Hal\Hbe \zeta-\q}{\zeta(\zeta-1)(\zeta-\a)}f = 0, \ee
with the condition
\be \label{Hcond} \Hga+\Hde+\Hep=\Hal+\Hbe+1, \qquad \a\ne 0,1. \ee
We use the upright type for the parameters $(\a, \q, \Hal, \Hbe, \Hga, \Hde, \Hep)$ for the Heun equation.
The Heun equation has four regular singular points at $\zeta=0,1,\a,\infty$, around which we can apply the Frobenius method to construct infinite power series solutions.
In particular, two local solutions at $\zeta=0$ are given by 
\begin{align}
\label{f01def} f_{01}(\zeta) &= Hl(\a, \q;\Hal, \Hbe, \Hga, \Hde;\zeta), \\
\label{f02def} f_{02}(\zeta) &= \zeta^{1-\Hga}Hl(\a, (\a\Hde+\Hep)(1-\Hga)+\q;\Hal+1-\Hga, \Hbe+1-\Hga, 2-\Hga, \Hde;\zeta),
\end{align}
whereas two local solutions at $\zeta=1$ are given by
\begin{align}
\label{f11def} f_{11}(\zeta) &= Hl(1-\a, \Hal\Hbe-\q;\Hal, \Hbe, \Hde, \Hga;1-\zeta), \\
\label{f12def} f_{12}(\zeta) &= (1-\zeta)^{1-\Hde}Hl(1-\a, ((1-\a)\Hga+\Hep)(1-\Hde)+\Hal\Hbe-\q;\Hal+1-\Hde, \Hbe+1-\Hde, 2-\Hde, \Hga;1-\zeta),
\end{align}
where $Hl(\a, \q;\Hal, \Hbe, \Hga, \Hde;\zeta)$ is known as the local Heun function or general Heun function HeunG, which converges for $|\zeta| < {\rm min}(1, |\a|)$.
The asymptotic behavior of the local solutions~\eqref{f01def}--\eqref{f12def} is governed by the characteristic exponents: 
\begin{align}
\label{asymy1} f_{01}(\zeta)&= 1 + \mO(\zeta), &  f_{02}(\zeta)&=  \zeta^{1-\Hga}[1+\mO(\zeta)], & (\zeta&\to 0) , \\
\label{asymy2} f_{11}(\zeta)&= 1 + \mO(1-\zeta), &  f_{12}(\zeta)&=  (1-\zeta)^{1-\Hde}[1+\mO(1-\zeta)], & (\zeta&\to 1) .
\end{align}

Inside the overlapping region of the disks of convergence, both local solutions at $\zeta=0,1$ are related to each other via linear combinations. 
Specifically, the local solutions at $z = 1$ can be written in terms of the local solution at $z = 0$ as
\begin{align}
f_{11}(z) = D_{11}f_{01}(z)+D_{12}f_{02}(z), \label{f11rel} \\ 
f_{12}(z) = D_{21}f_{01}(z)+D_{22}f_{02}(z), \label{f12rel}  
\end{align}
where the connection coefficients are given by
\be \label{D11} D_{11}=\f{W_\zeta[f_{11},f_{02}]}{W_\zeta[f_{01},f_{02}]}, \qquad
D_{12}=\f{W_\zeta[f_{11},f_{01}]}{W_\zeta[f_{02},f_{01}]}, \qquad 
D_{21}=\f{W_\zeta[f_{12},f_{02}]}{W_\zeta[f_{01},f_{02}]}, \qquad
D_{22}=\f{W_\zeta[f_{12},f_{01}]}{W_\zeta[f_{02},f_{01}]}, \ee
and $W_\zeta[u,v] = u \f{dv}{d\zeta} - \f{du}{d\zeta} v$ is the Wronskian with differentiation with respect to $\zeta$.

\subsection{Angular solution}%%%%%%%%%%%%%%%%%%%%%

We are now ready to rewrite the angular function as 
\be S_s(\theta)=x^{A_1}(x-1)^{A_2}(x-x_\a)^{A_3}(x-x_\infty) w_{s}(x) , \ee
where
\begin{align}
x&= \f{(1-i/\sqrt{\lamas})(\mu+1)}{2(\mu-i/\sqrt{\lamas})}, \qquad 
x_\infty=\f{1-i/\sqrt{\lamas}}{2}, \qquad
x_\a = -\f{(1-i/\sqrt{\lamas})^2}{4i/\sqrt{\lamas}}, \\
A_1&=\f{m-s}{2}, \quad
A_2=-\f{m+s}{2}, \quad  
A_3=\f{1}{2}\left[ s + i\mk{ \f{1+\lamas}{\sqrt{\lamas}}c -m \sqrt{\lamas} } \right], \quad 
A_4=\f{1}{2}\left[ s - i\mk{ \f{1+\lamas}{\sqrt{\lamas}}c -m \sqrt{\lamas} } \right],
\end{align}
and then $w_s(x)$ obeys the general Heun equation~\eqref{Heun} with 
\begin{align} 
\label{param-ang} 
\zeta&=x, \qquad 
f= w_s, \qquad  
\a=x_\a, \qquad 
\q=\f{i\lambda}{4\sqrt{\lamas}}+\f{1}{2}+A_1 +\left(m+\f{1}{2}\right)(A_3 -A_4),\notag\\
\Hal&=1,\qquad
\Hbe=-2 A_4+1,\qquad
\Hga=2A_1+1,\qquad
\Hde=2A_2+1,\qquad
\Hep=2A_3+1. 
\end{align}
The local solutions at $x=0,1$ are given by $w_{01,s}(x), w_{02,s}(x)$ and $w_{11,s}(x), w_{12,s}(x)$, corresponding to \eqref{f01def}--\eqref{f12def}.

Given the asymptotic behaviors~\eqref{asymy1} and \eqref{asymy2}, for $S_s(\theta)$ to be regular at $\theta=0,\pi$, we require the following linear dependence of the exact solutions:
\be \label{exactlambda}
W_x[w_{0i,s},w_{1j,s}]=0,\qquad
i=\begin{cases}
1 & (m-s\geq 0) \\
2 & (m-s< 0) ,
\end{cases} \qquad
j=\begin{cases}
1 & (m+s\leq 0) \\
2 & (m+s> 0) .
\end{cases} 
\ee
This requirement determines the separation constant $\lambda$.
Since the left-hand side of \eqref{exactlambda} involves $\lambda$ in a nontrivial way, one may use a root-finding algorithm to determine $\lambda$.

\subsection{Radial solution}%%%%%%%%%%%%%%%%%%%%%

Similarly, we rewrite the radial function as 
\be R_s(r)=z^{B_1} (z - 1)^{B_2} (z - z_\a)^{B_3} (z - z_\infty)^{2 s + 1} y_s(z) \ee
where 
\begin{align} 
z&=\f{r'_+-r_-}{r'_+-r_+} \f{r-r_+}{r-r_-}, \qquad 
z_\infty=\f{r'_+-r_-}{r'_+-r_+}, \qquad
z_\a = z_\infty\f{r'_--r_+}{r'_--r_-}, \\
B_1&=B(r_+), \qquad
B_2=B(r'_+), \qquad 
B_3=B(r'_-), \qquad 
B_4=B(r_-),
\end{align}
with
\be \label{Bdef} B(r) 
= \f{i\KJ(r)}{\Delta'(r)} 
= \f{i}{2\kappa(r)}\mk{\omega - m \f{a}{r^2+a^2}}, \ee
and then $y_s(z)$ obeys the general Heun equation~\eqref{Heun} with 
\begin{align} 
\label{param-rad} \zeta&=x, \qquad 
f=y_s, \qquad  
\a=z_\a, \notag\\
\q&=-\f{ \lambda - 2 s (1 - \lamas) - \f{\Lambda}{3} (s+1)(2s+1)(r_+ r_- + r'_+ r'_-) }{\f{\Lambda}{3} (r_- - r'_-) (r_+ - r'_+) } +\f{ 2 i (2s+1) (1 + \lamas) [ \omega (r_+ r_- + a^2) - a m ] }{\f{\Lambda}{3} (r_- - r'_-) (r_- - r_+) (r_+ - r'_+)} ,\notag\\
\Hal&=2s+1,\qquad
\Hbe=-2B_4+s+1,\qquad
\Hga=2B_1+s+1,\qquad
\Hde=2B_2+s+1,\qquad
\Hep=2B_3+s+1. 
\end{align}
Again, the local solutions at $z=0,1$ are given by $y_{01,s}(z), y_{02,s}(z)$ and $y_{11,s}(z), y_{12,s}(z)$, in parallel to \eqref{f01def}--\eqref{f12def}.
The corresponding radial solutions are denoted by $R_{01,s}(r), R_{02,s}(r)$ and $R_{11,s}(r), R_{12,s}(r)$.

From the asymptotic behavior~\eqref{asymy1} and \eqref{asymy2}, we obtain
\begin{align}
\label{R-asymy1} R_{01,s}(r)&\propto (r-r_+)^{-s/2+\theta_+}, &  R_{02,s}(r)&\propto (r-r_+)^{-s/2-\theta_+}, & (r&\to r_+), \\
\label{R-asymy2} R_{11,s}(r)&\propto (r-r'_+)^{-s/2+\theta_\c}, &  R_{12,s}(r)&\propto (r-r'_+)^{-s/2-\theta_\c}, & (r&\to r'_+) ,
\end{align} 
where
\be \theta_+ = B(r_+) + \f{s}{2}, \qquad \theta_\c = B(r'_+) + \f{s}{2} . \ee
By definitions~\eqref{rstar} and \eqref{Bdef}, we see that $(r-r_h)^{\pm \theta_h} \sim e^{\pm i\omega r^*}$ at the vicinity of the horizon $r\sim r_h$, which, respectively, describes outgoing and ingoing waves in the tortoise coordinates.

\section{Graybody factor}%%%%%%%%%%%%%%%%%%%%%%%%%%%%%%%%%%%%%%%%%
In this section, we consider a scattering problem of a Kerr-de Sitter black hole with a reflective surface located slightly outside the outer horizon radius. We then compute the graybody factor of the black hole.

\subsection{Classical black hole}%%%%%%%%%%%%%%%%%%%%%

For the standard case, we define the ``up'' solution as a purely outgoing wave at $r\to r'_+$, i.e.,
\begin{align}
\label{Rup-asym-BH} R_{{\rm up},s}(r) &\to
\begin{cases}
\displaystyle (r-r_+)^{-s/2+ \theta_+} - \mR_s (r-r_+)^{-s/2- \theta_+} & (r\to r_+)\\
\displaystyle \mT_s (r-r'_+)^{-s/2+ \theta_\c} & (r\to r'_+) .
\end{cases}
\end{align}
The coefficients $\mR_s$ and $\mT_s$ amount to the reflection and transmission coefficients of the angular momentum barrier for the spin-$s$ field, respectively. 
From the conservation of $\Delta^{s + 1} W_r[R_{{\rm up},s},R_{{\rm down},s}]$, where $R_{{\rm down},s}(r)=\Delta^{-s}R_{{\rm up},-s}^*(r)$, we obtain 
\be \mR_s \mR_{-s}^* + F_s^{-1} \mT_s \mT_{-s}^* = 1,  \ee
where
\be F_s=\f{\theta_+ \Delta'(r_+)}{\theta_\c \Delta'(r_\c)} . \ee
Thus, the graybody factor~$\Gamma_{slm}^{\rm (O)}(\omega)$ is defined as 
\be \label{grey-def-BH} \Gamma_{slm}^{\rm (O)} \equiv F_s^{-1} \mT_s \mT_{-s}^* 
. \ee

With the general Heun functions, we can write down $R_{{\rm up},s}(r)$ exactly.
From the asymptotic behavior~\eqref{R-asymy1} and \eqref{R-asymy2} and the connection relation~\eqref{f11rel}, we can identify  
\begin{align} 
\label{Rup-Heun} R_{{\rm up},s}(r) &= \begin{cases}
D_{11,s}R_{01,s}(r)+D_{12,s}R_{02,s}(r) & (r\to r_+)\\
R_{11,s}(r) & (r\to r'_+) ,
\end{cases}
\end{align} 
up to an overall factor.
Comparing the asymptotic form of \eqref{Rup-Heun} with \eqref{Rup-asym-BH}, and using \eqref{D11}, we can write down the reflection and transmission coefficients as
\begin{align}
\mR_s &= \f{W_z[y_{11,s},y_{01,s}]}{W_z[y_{11,s},y_{02,s}]} \mk{ \f{r_+-r_-}{z_\infty} }^{2B_1+s} , \\
\mT_s &= \f{W_z[y_{01,s},y_{02,s}]}{W_z[y_{11,s},y_{02,s}]} \mk{ \f{z_\infty-1}{z_\infty} }^{2s+1} 
\mk{ \f{r_+-r_-}{z_\infty} }^{B_1} \mk{ \f{z_\infty(r_+-r_-)}{-(r'_+-r_-)^2} }^{B_2} \mk{ \f{z_\a-1}{z_\a} }^{B_3}  . 
\end{align}
We then obtain the graybody factor~\eqref{grey-def-BH} as 
\be \label{grey-BH} \Gamma_{slm}^{\rm (O)} = F_s^{-1} \mk{\f{z_\infty-1}{z_\infty}}^2 \f{W_z[y_{01,s},y_{02,s}]}{W_z[y_{11,s},y_{02,s}]} \mk{ \f{W_z[y_{01,-s},y_{02,-s}]}{W_z[y_{11,-s},y_{02,-s}]} }^*  . \ee

\subsection{Black hole with a reflective surface}%%%%%%%%%%%%%%%%%%%%%
\begin{figure}[t]
  \begin{center}
    \includegraphics[keepaspectratio=true,height=100mm]{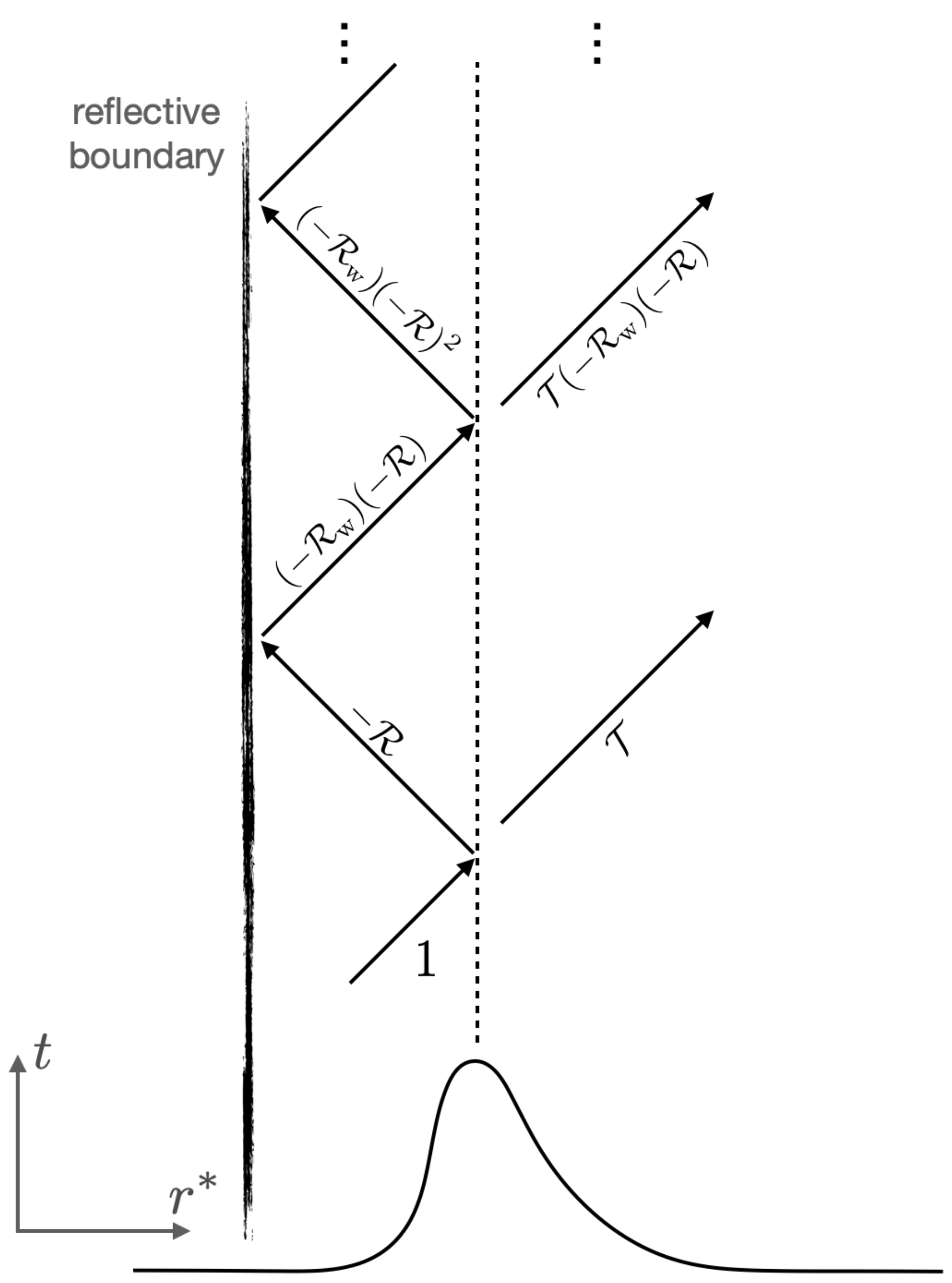}
  \end{center}
\caption{Schematic picture showing the resonance in the cavity between the reflective surface and angular momentum barrier. The reflective boundary condition modifies the graybody factor.
}
\label{schematic}
\end{figure}

In addition to the angular momentum barrier located at $r^{\ast} \sim \mO(M)$, we introduce a reflective surface of a black hole, which we assume is located slightly outside the outer horizon radius.
Let $r^{\ast} = r^{\ast}_{\rm w}$ be the location of the reflective surface in tortoise coordinates, and $\mR_{{\rm w},s}$ be the reflection coefficient of the reflective surface for the spin-$s$ field. 
Similarly to the standard case, we define the up solution $R_{{\rm up},s}(r)$ for the radial Teukolsky equation as a solution satisfying a purely outgoing boundary condition at $r=r_+'$. Summing up the infinite scattering between the reflective surface and the angular momentum barrier (see Fig. \ref{schematic}), $R_{{\rm up},s}(r)$ can be written in terms of the transmission and reflection coefficients as
\begin{align}
\label{Rup-asym} R_{{\rm up},s}(r) &\to
\begin{cases}
\displaystyle \f{1}{1- \mR_s \mR_{{\rm w},s} } [ (r-r_+)^{-s/2+ \theta_+}  -\mR_s (r-r_+)^{-s/2- \theta_+} ] & (r^{\ast}_{{\rm w}} \leq r^{\ast} \ll -M)\\
\displaystyle \f{ \mT_s }{1- \mR_s \mR_{{\rm w},s} } (r-r_+')^{-s/2+ \theta_{\rm c}} & (r^{\ast} \gg M) .
\end{cases}
\end{align}
The graybody factor $\Gamma_{slm}(\omega)$ is then given by
\begin{equation}
\Gamma_{slm}(\omega) \equiv F_s^{-1} \f{\mT_s}{1-\mR_s \mR_{{\rm w},s}} \mk{\f{\mT_{-s}}{1-\mR_{-s} \mR_{{\rm w},-s}}}^*.
\end{equation}
Note that the limit $\mR_{{\rm w},s}\to 0$ recovers the up mode~\eqref{Rup-asym-BH} and the graybody factor~\eqref{grey-def-BH} without the reflective surface.

%%%%%%%%%%%%%%%%%%%%%%%%%
%%%%%%%%%%%%%%%%%%%%%%%%%

The reflectivity $\mR_{{\rm w},s}$ at the reflective surface is modeled by its absolute value $\epsilon_s$ and phase shift $\tilde{\delta}_s$. Also, the position of the wall, $r^{\ast} = r^{\ast}_{{\rm w},s}$, determines the phase delay. Taking into account these factors, we have
\begin{equation}
\mR_{{\rm w},s} = \epsilon_s \exp \left[ -2ir^{\ast}_{{\rm w},s} k_{\rm H} + i \tilde{\delta}_s \right],
\label{wall_reflection_co}
\end{equation}
where $k_{\rm H} \equiv \omega - m \Omega_{\rm H}$ with $\Omega_{\rm H} \equiv a/(r_+^2 + a^2)$.
The factor of $-2i r^{\ast}_{{\rm w},s} k_{\rm H}$ in (\ref{wall_reflection_co}) is the phase delay caused by the optical path difference of the cavity between the reflective surface at $r^{\ast} = r^{\ast}_{{\rm w},s}$ and angular momentum barrier at $r^{\ast} \sim 0$.
In order for the graybody factor to be real, here we impose
\begin{equation}
\mR_s \mR_{{\rm w},s} = \mR_{-s} \mR_{{\rm w},-s}.
\label{real_gray_condition}
\end{equation}
In the Chandrasekhar-Detweiler variable, this condition corresponds to merely requiring that the reflection coefficient in energy is real \cite{Oshita:2020dox}. Depending on the perturbation variable we choose, the condition in (\ref{real_gray_condition}) can be either trivial or complicated.
Then the graybody factor reduces to
\begin{equation} \label{grey-ref}
\Gamma_{slm} (\omega) = \frac{\Gamma_{slm}^{\rm (O)}}{1+|\mR_s|^2 \epsilon_s^2 -2 |\mR_s| \epsilon_s \cos{\left[ \Theta_s -2 r^{\ast}_{{\rm w},s} k_{\rm H} \right]}},
\end{equation}
where $\Theta_s \equiv \delta_s + \tilde{\delta}_{s}$ and $\delta_s \equiv \text{Arg} (\mR_s)$. Again, we see that if the black hole horizon is a perfect absorber ($\epsilon_s=0$), the graybody factor reduces to the original one. In the following, we will omit the subscript $s$ for the spin when it does not cause ambiguity.

\begin{figure}[t]
  \begin{center}
    \includegraphics[keepaspectratio=true,height=50mm]{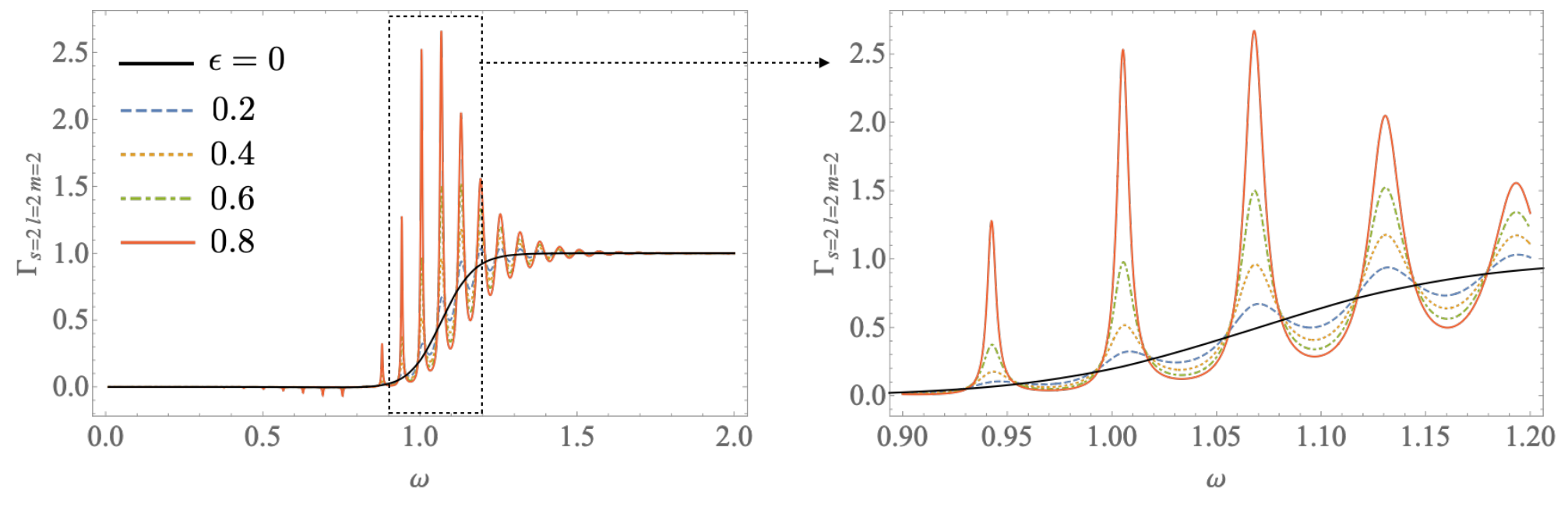}
  \end{center}
\caption{Graybody factor for various values of $\epsilon$. Here we set other parameters as $M = 1/2$, $a/M=0.7$, $\Lambda = 5 \times 10^{-4}$, $(s,l,m)=(2,2,2)$, $r^{\ast}_{\rm w} = -50$, and $\Theta = 0$.
}
\label{greybody_1}
%%%%%%%%%%%%%%%%%%%
\begin{center}
    \includegraphics[keepaspectratio=true,height=40mm]{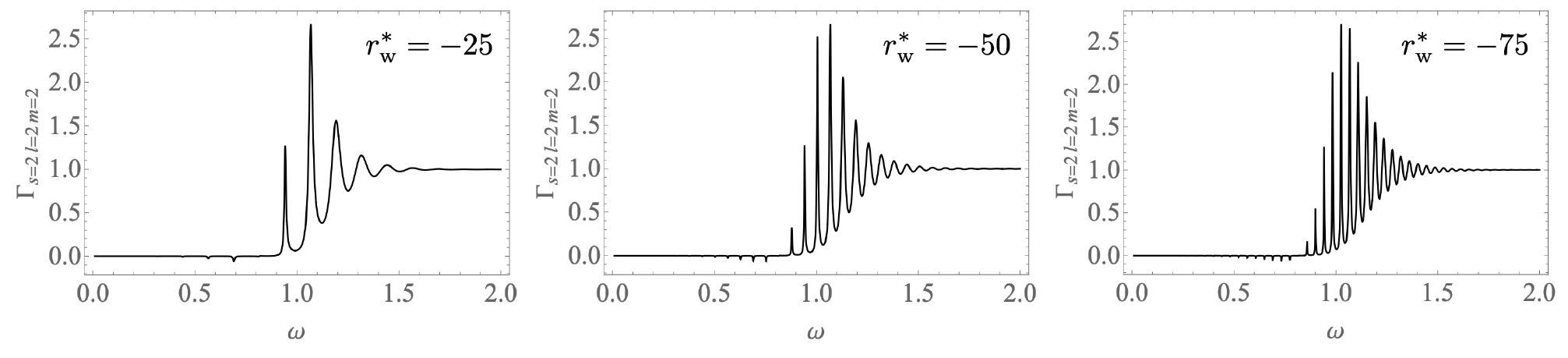}
  \end{center}
\caption{Graybody factor for various values of $r^{\ast}_{\rm w}$, where $M = 1/2$, $a/M=0.7$, $\Lambda = 5 \times 10^{-4}$, $\epsilon = 0.8$, $(s,l,m)=(2,2,2)$, and $\Theta = 0$. 
}
\label{greybody_2}
\end{figure}

In the presence of the reflective surface with $\epsilon > 0$, we expect that a signature of the resonance in the region of $r^{\ast}_{\rm w} \leq r^{\ast} \lesssim M$ shows up in the graybody factor.
Indeed, as shown in Fig.~\ref{greybody_1}, the graybody factor is subject to a high-frequency modulation in the frequency space.
As expected, the amplitude of the modulation increases as $\epsilon$ increases.
The resonance frequency is determined by the distance between the angular momentum barrier and the reflective surface and hence 
depends on $r^{\ast}_{\rm w}$.
In Fig.~\ref{greybody_2}, we see that the density of resonant peaks in the frequency domain is roughly proportional to $|r^{\ast}_{\rm w}|$. This behavior is consistent with the factor $\cos{\left[ \Theta -2 r^{\ast}_{\rm w} (\omega - m \Omega_{\rm H}) \right]}$ in \eqref{grey-ref}.

The values of the parameters $\epsilon$, $r^{\ast}_{\rm w}$, and $\Theta$ are model dependent, and there is no promising theoretical model to fix them at least at this moment. Therefore, in the next section, we comprehensively explore possible parameter regions to investigate the effect of the horizon reflectivity on the black hole evaporation. Also, it is important to note that the assumption of nearly perfect reflectivity may put an upper constraint on the lifetime of an echoing black hole.
%%%%%%%%%%%%%%%%%%%%%%%%%
%%%%%%%%%%%%%%%%%%%%%%%%%

\section{Evaporation process of an echoing black hole}%%%%%
Here we numerically compute the time evolution of the mass and angular momentum of a black hole with the reflective surface. The near-horizon structure of the black hole is assumed to be thermal and reflective due to quantum gravitational effects. The reflective boundary condition near the horizon leads to the resonance (echoes) as shown in the previous section. In this section, we investigate how the lifetime of the black hole emitting thermal radiation is affected by the echo mechanism.

A technical caveat is that our procedure does not work for the (asymptotically flat) Kerr background, for which the master equation takes the form of the confluent Heun equation, instead of the Heun equation for the Kerr-de Sitter background.
In this case, one needs to deal with the connection problem between the solutions at the regular singular point (black hole horizon) and the irregular singular point (infinity), which is more complicated than the case of the Heun equation.
To circumvent this difficulty, one can make use of the extrapolation of the results for small cosmological constant to predict high-precision results for the Kerr background~\cite{Hatsuda:2020sbn}. 
Nevertheless, to clarify qualitative behaviors, we can still rely on the results for the Kerr-de Sitter background with a small cosmological constant without extrapolation.
Actually, with a sufficiently small cosmological constant, numerical results are quite stable unless one requires high numerical precision.
Thus, in the following, we present the results with a small cosmological constant\footnote{A similar issue arises in the extremal case, for which the Heun's differential equation has an irregular singular point at the black hole horizon. It makes the extension of our computation to the extremal case difficult.}.
In the Appendix, we provide a consistency check that the small cosmological constant does not affect our results, as well as a resolution test for our numerical calculation.

\subsection{Mass-loss and spin-loss rates}
\begin{figure}[h]
  \begin{center}
    \includegraphics[keepaspectratio=true,height=40mm]{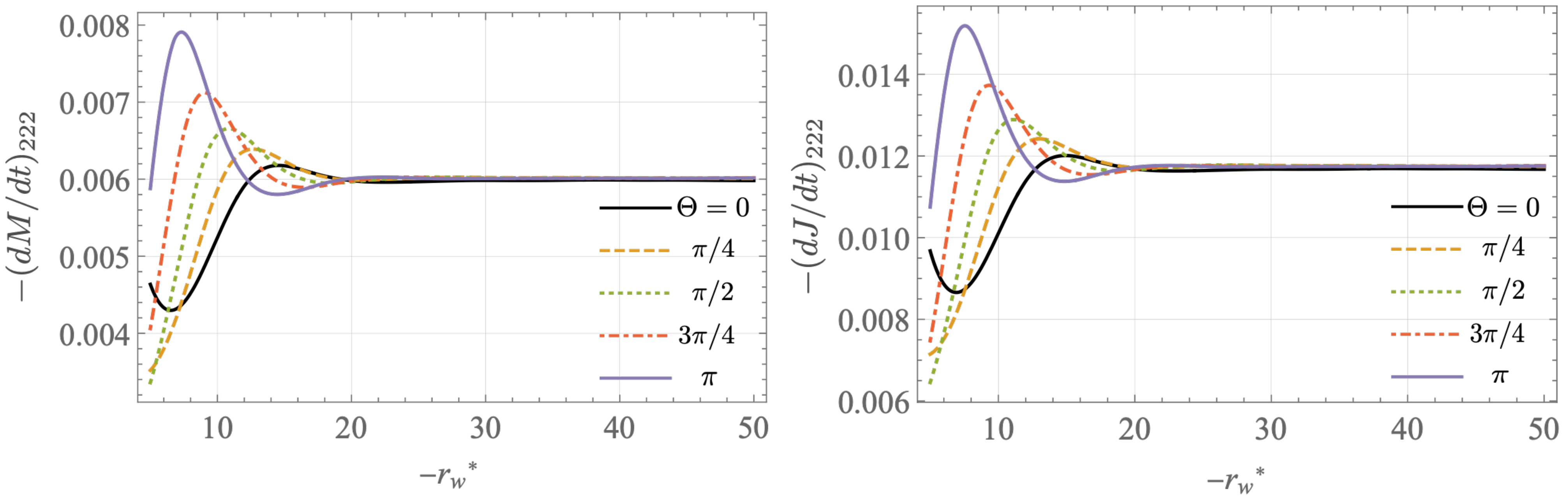}
  \end{center}
\caption{The $r^{\ast}_{\rm w}$ dependence of the mass- and spin-loss rates for various values of $\Theta$. The other parameters are set to $M = 1/2$, $a/M=0.7$, $\Lambda = 5 \times 10^{-4}$, $\epsilon=0.8$, and $(s,l,m) = (2,2,2)$.
}
\label{flux_delta}
  \begin{center}
    \includegraphics[keepaspectratio=true,height=40mm]{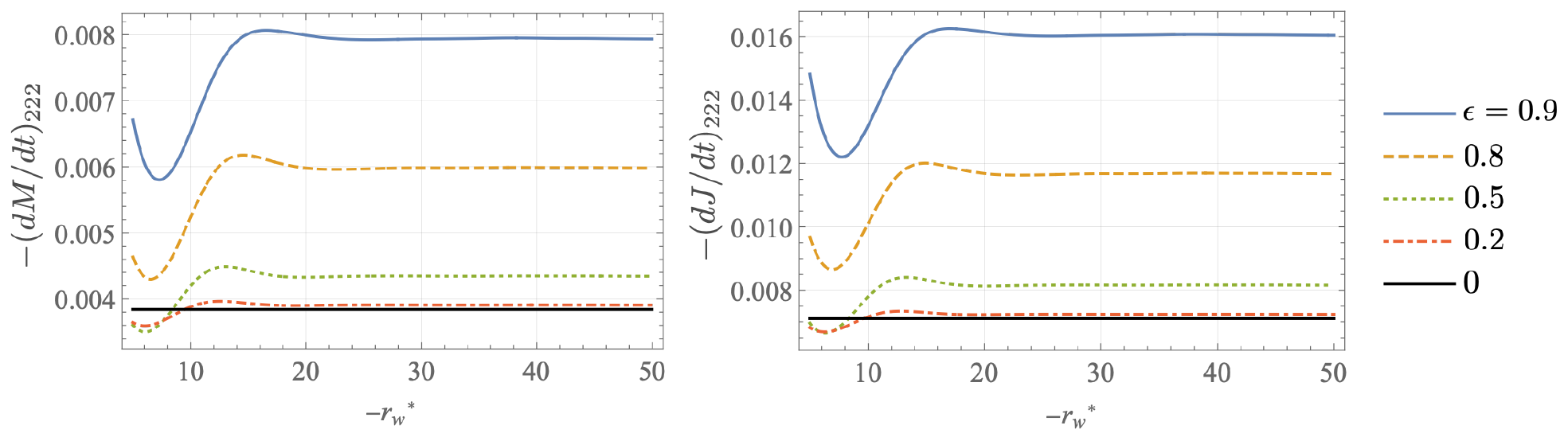}
  \end{center}
\caption{The $r^{\ast}_{\rm w}$ dependence of the mass- and spin-loss rates for various values of $\epsilon$. The other parameters are set to $M = 1/2$, $a/M=0.7$, $\Lambda = 5 \times 10^{-4}$, $\Theta = 0$, and $(s,l,m) = (2,2,2)$.
}
\label{flux_eps}
\end{figure}

In this subsection, we show the dependence of the mass-loss and spin-loss rates of a rotating black hole with the reflective boundary condition controlled by the parameter set $(\epsilon, r^{\ast}_{\rm w}, \Theta)$. Here we assume that echoing black holes emit thermal radiation that has the blackbody spectrum with the Hawking temperature. Also, for simplicity, the parameters characterizing the reflective surface are assumed to be independent of the species of emitted particles. The mass- and spin-loss rates are given by~\cite{Page:1976ki}
\begin{align}
\frac{dM}{dt} & = -\sum_{slm} \frac{1}{2 \pi} \int^{\infty}_{0}  d\omega \frac{\omega \Gamma_{slm} (\omega,\epsilon, r^{\ast}_{\rm w}, \Theta)}{e^{k_{\rm H}/T_{\rm H}} -(-1)^{2s}}= \sum_{slm} \left(\frac{dM}{dt} \right)_{slm},\\
\frac{dJ}{dt} &= -\sum_{slm} \frac{1}{2 \pi} \int^{\infty}_{0} d\omega \frac{m \Gamma_{slm} (\omega,\epsilon, r^{\ast}_{\rm w}, \Theta)}{e^{k_{\rm H}/T_{\rm H}} -(-1)^{2s}}= \sum_{slm} \left(\frac{dJ}{dt} \right)_{slm},
\end{align}
where $J\equiv a M$ is the magnitude of the angular momentum for the Kerr black hole.
Although the power spectrum depends on the three parameters $(\epsilon, r^{\ast}_{\rm w}, \Theta)$, the fluxes $dM/dt$ and $dJ/dt$ are insensitive to the two parameters $r^{\ast}_{\rm w}$ and $\Theta$ when $|r^{\ast}_{\rm w}| \gg 1$ as is shown in Fig. \ref{flux_delta} and \ref{flux_eps}. In those figures, we show the mass-loss and spin-loss rates associated with $(s,l,m) = (2,2,2)$. The reason why the evaporation rate is insensitive to $r^{\ast}_{\rm w}$ and $\Theta$ for $|r^{\ast}_{\rm w}| \gg 1$ is that the frequency integration involving the {\it dense} resonant peaks of the graybody factor coarse grains the resonant fine structure (see Fig. \ref{greybody_2}). 
Therefore, if the reflective surface is sufficiently close to the outer horizon, the contribution of the reflective surface to the evaporation process is solely governed by the reflectivity parameter $\epsilon$.
In the following, we consider such a case and fix the two parameters as $r^{\ast}_{\rm w} = -50$ and $\Theta = 0$ without loss of generality.
%%%%%%%%%%%%%%%%%%%%%%%%%
%%%%%%%%%%%%%%%%%%%%%%%%%
%%%%%%%%%%%%%%%%%%%%%%%%%
\begin{figure}[h]
  \begin{center}
    \includegraphics[keepaspectratio=true,height=65mm]{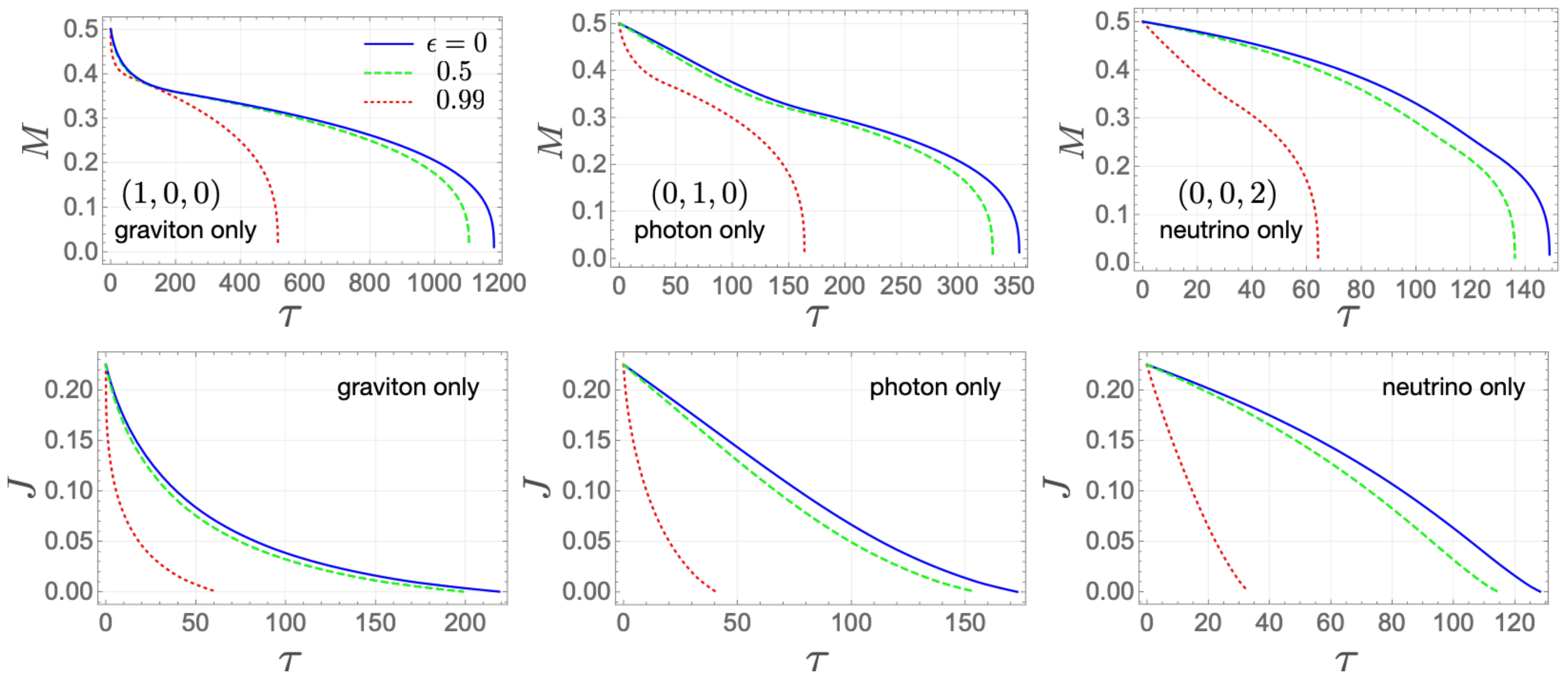}
  \end{center}
\caption{Time development of the mass and angular momentum of a black hole. The reflectivity is set to $\epsilon = 0$ (blue solid), $0.5$ (green dashed), and $0.99$ (red dotted). The particle species are set to $(n_2, n_1, n_{1/2}) = (1,0,0)$ (left), $(0,1,0)$ (center), and $(0,0,2)$ (right). 
}
\label{timeD}
\end{figure}
%%%%%%%%%%%%%%%%%%%

\subsection{Time evolution of an echoing black hole}
\label{sec:sub:time_evol}
We numerically investigate the time development of the mass and angular momentum of an echoing black hole whose mass-loss and spin-loss rates are controlled only by the reflectivity parameter $\epsilon$ when $|r^{\ast}_{\rm w}| \gg 1$. The initial mass and spin of the black hole are set to $M_o \equiv M (t=0) = 1/2$ and $j(t=0) = 0.9$, where $j\equiv a/M$ is a dimensionless spin parameter. 
The number of spin-$s$ particle species emitted from the black hole is labeled by $n_{s}$. 
To follow the development of $z \equiv -\log (M/M_{o})$ with respect to $y \equiv \log j$, as was performed in Ref.~\cite{Page:1976ki}, we numerically solve the following derivative equations with the 4th-order Runge-Kutta method:
\begin{align}
\frac{dz}{dy} &= \frac{f}{g-2f},
\label{z_eq}\\
\frac{d \tau}{dy} &= \frac{e^{-3z}}{g-2f},
\label{tau_eq}
\end{align}
where 
\begin{align}
\tau &\equiv M_o^{-3} t,\\
f &\equiv -M^3 \frac{d (\ln M)}{dt},\label{f_def}\\
g &\equiv -M^3 \frac{d (\ln J)}{dt}.
\end{align}
We assume that the spin parameter $j(t)$ is a monotonic function when solving the differential equations (\ref{z_eq}) and (\ref{tau_eq}) to predict the time development of $M(t)$ and $J(t)$. The numerical computation is truncated at $j = 0.0009$. For $j \leq 0.0009$, the time development of $M(t)$ is computed simply by solving (\ref{f_def}) by approximating $f$ as a constant. In other words, we neglect the rotation effect in the evaporation process for $j \leq 0.0009$. 

Figure \ref{timeD} shows the time development of the total mass and angular momentum of a black hole that emits only gravitons, photons, and neutrinos, respectively. We take into account dominant harmonic modes for each species: $(l,m) =(2,2)$ and $(3,3)$ for gravitons, $(l,m) = (1,1)$ and $(2,2)$ for photons, and $(l,m) = (1/2,1/2)$ and $(3/2,3/2)$ for neutrinos. This shows that the lifetime of an echoing black hole that spontaneously emits thermal radiation is shortened by a factor of $\sim 2$ for a nearly perfect reflectivity $\epsilon = 0.99$. Note that to estimate a lower bound on the lifetime of the evaporating black hole, here we consider a nearly-perfect reflectivity as an extreme situation, although it is nontrivial if the thermalization of the reflective surface and its nearly perfect reflectivity can be compatible.
As the evaporation proceeds and the angular momentum is extracted by Hawking particles, the symmetry of the background spacetime approaches the spherical symmetry, and lower angular modes contribute to the dominant energy flux of the Hawking radiation (see, e.g., Fig. 7 in \cite{Gregory:2021ozs}). As such, the lower-spin fields mainly contribute to the emission of Hawking particles after the angular momentum of the black hole is sufficiently extracted.

\begin{figure}[t]
  \begin{center}
    \includegraphics[keepaspectratio=true,height=75mm]{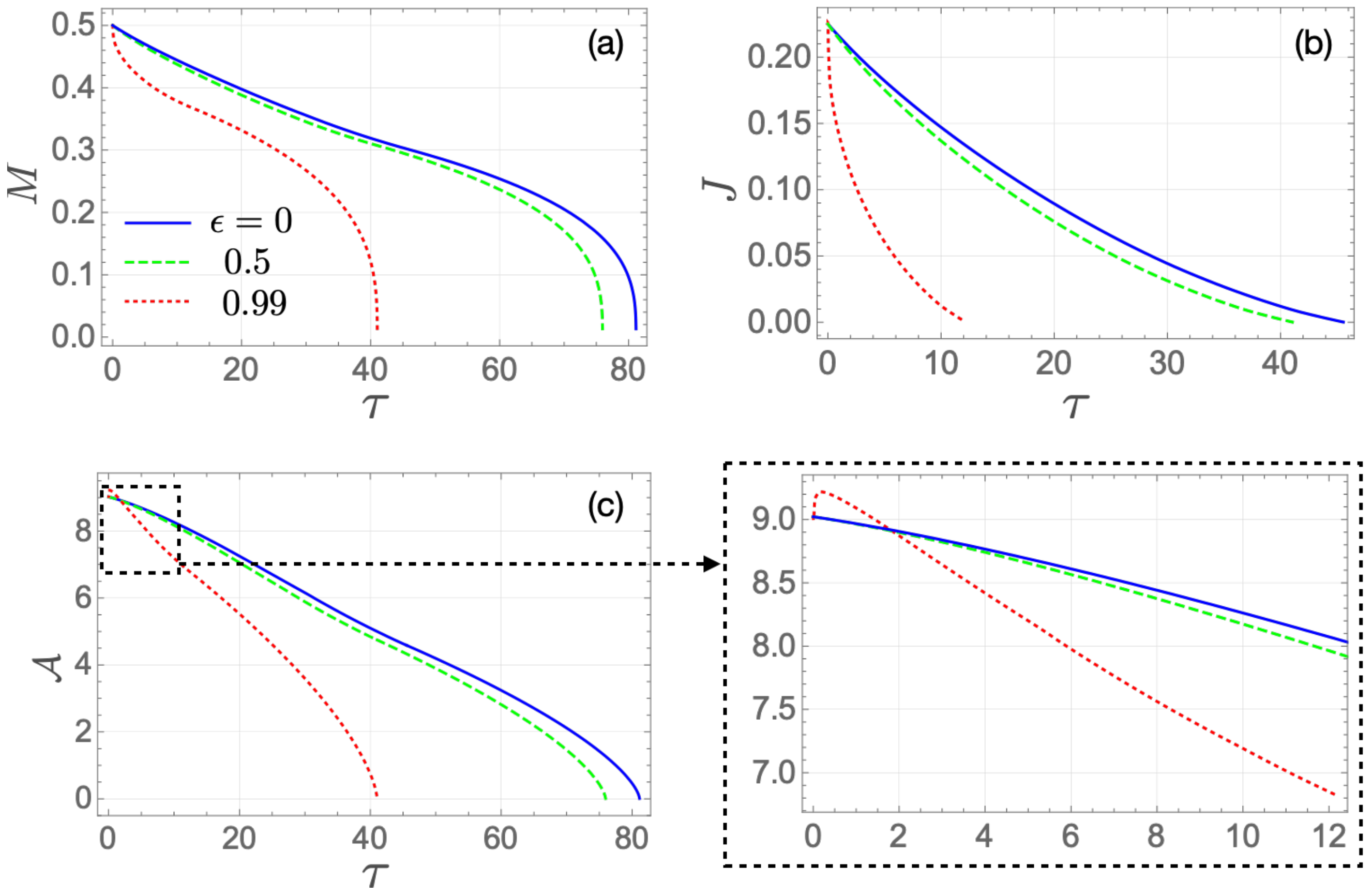}
  \end{center}
\caption{(a) Mass and (b) angular momentum of a black hole with the same parameter set as in Fig. \ref{timeD}. The particle species are set to $(n_2, n_1, n_{1/2}) = (1,1,2)$. The area of the black hole horizon is also shown in (c), and it can be seen that the area increases when $j\sim 0.9$ due to the rapid decay of its angular momentum.
}
\label{all_timeD}
\end{figure}

On the other hand, Fig. \ref{all_timeD}(a) and \ref{all_timeD}(b) show $M=M(\tau)$ and $J=J(\tau)$ for a black hole that emits all particles with $(n_2,n_1,n_{1/2}) = (1,1,2)$. Here we take a minimal combination of particle species: $n_2=1$ for gravitons, $n_1=1$ for photons, and $n_{1/2} =2$ for electron neutrinos and muon neutrinos. We also take into account the 2 degrees of freedom of polarization or helicity for each species.
While the lifetime itself of the echoing black hole varies depending on how many particle species are taken into account for the evaporation procedure, it is roughly reduced by several factors for a nearly perfect reflectivity $\epsilon = 0.99$. 

Let us consider the time evolution of the horizon area ${\mathcal A}$, which has the form
\begin{equation}
{\mathcal A} = 4 \pi (r_+^2 +a^2) = 2 \pi M r_+.
\end{equation}
In Fig. \ref{all_timeD}(c), one can see that the area monotonically decreases for a nonextremal black hole\footnote{In Ref.~\cite{Page:1976ki}, the author shows that the area increases at the near-extremal limit and then decreases until the hole evaporates.} with its small reflectivity. On the other hand, the area initially increases for a rapidly spinning black hole ($j \sim 0.9$) with its nearly perfect reflectivity. The instantaneous increment of the horizon area may be caused by superradiance enhanced by the reflective surface in the ergosphere. The amplified superradiance leads to the sudden increase of $r_+ = M + \sqrt{M^2-a^2}+\mO(\Lambda M^3)$ as it is a monotonically decreasing function with respect to the spin parameter $a$. In Ref.~\cite{Gregory:2021ozs}, it was reported that a positive cosmological constant suppresses the maximum value of the superradiant amplification, $\max (- \Gamma_{slm}^{(O)})$. Based on the result, the increase of the horizon area at rapid rotations might be suppressed by a positive cosmological constant. It is interesting to test the expectation and to extend it to the case of a negative cosmological constant, but this is not our focus here.
%%%%%%%%%%%%%%%%%%%%%%%%%
%%%%%%%%%%%%%%%%%%%

\section{Conclusion}%%%%%%%%%%%%%%%%%%%%%%%%%%%%%%%%%%%%%%%%%
We considered an {\it echoing} black hole that has a reflective surface slightly outside the radius of the outer horizon such as a stretched horizon or firewall. Assuming the spontaneous emission of thermal radiation from the reflective surface, we have shown that the resonance of Hawking radiation in the cavity formed between the reflective surface and angular momentum barrier results in the enhancement of the flux carrying the mass and angular momentum of the black hole. 

The reflective surface is assumed to be characterized by three factors: constant reflectivity, phase shift, and position. We first computed the graybody factor by using the exact solutions of the Teukolsky equation that can be transformed to the Heun's differential equation. 
We have found that the graybody factor has resonant modulations in the frequency domain and that the resonant frequency depends on the phase shift and position of the surface. Nevertheless, we have shown that the energy fluxes carrying the mass and angular momentum of the black hole are insensitive to those two factors in the close limit of the reflective surface ($r^{\ast}_{\rm w} \ll -1$).

We have performed the numerical computation to follow the time development of the mass and angular momentum of the black hole by taking into account the emission of gravitons, photons, and neutrinos. We have found that the reflective surface promotes the evaporation process and that the lifetime is shorten by several factors for a nearly perfect reflectivity ($\epsilon = 0.99$). 
This means that the reflective surface of a quantum black hole not only changes the gravitational wave signals at late times but also affects its evaporation process and lifetime. 
We have also found that the superradiance is enhanced by the reflective surface (i.e., ergoregion instability) and that the area of a highly spinning black hole increases due to the rapid decay of its spin even for $j \leq 0.9$. In the previous work by Page \cite{Page:1976ki}, a similar phenomenon was reported for a near-extremal black hole without the reflective surface.

Although the reflectivity of black holes is a completely unknown factor, an actual reflectivity might be milder than the $\epsilon=0.99$ we assumed in our computation since the stretched horizon is thought to involve dissipative effects \cite{Susskind:1993if}. Also, another model, the quantized black hole horizon, would lead to the mixture of perfect absorption and reflection depending on frequency \cite{Cardoso:2019apo}. 
In the Boltzmann-reflectivity model \cite{Oshita:2018fqu,Oshita:2019sat,Wang:2019rcf}, the reflectivity $\epsilon^2 = e^{-|k_{\rm H}|/T_{\rm H}}$ is determined by the thermal nature of the black hole, i.e., the Hawking temperature $T_{\rm H}$ and the angular velocity of the horizon (or a chemical potential) $m\Omega_{\rm H}$. If this is the case, only the modes of $\omega \sim m \Omega_{\rm H}$ contribute to the echo evaporation, and other modes are mostly dissipated at the surface.
Therefore, $\epsilon=0.99$ is a stronger assumption, and the lifetime obtained from it may be a lower bound of the lifetime of an actual quantum black hole.
Our conclusion is that if evaporating black holes have nontrivial surfaces, such as firewalls, stretched horizons and so on, and have nonzero reflectivity as has mainly been discussed in the context of gravitational-wave echoes, the lifetime of black holes may be shorter than the standard lifetime by a factor of at most $\sim 2$ for $\epsilon \leq 0.99$ and $j \leq 0.9$. In other words, we found that in most cases, there is no change of the order of magnitude in the lifetime. The existing cosmological constraint on the mass of primordial (microscopic) black holes would therefore be insensitive to the scenario of black hole information recovery.

As possible extensions of our model, we could introduce the spin-$s$ dependence or frequency dependence of the reflective surface. As a specific example where the frequency dependence can be important, one may consider an evaporation process of a wormhole if it emits quantum radiation. Its geometry could be modeled by the junction of two Schwarzschild solutions as was demonstrated in Ref.~\cite{Cardoso:2016rao}. In that case, the echo of Hawking flux is significant only for low-frequency modes, which can be modeled by introducing the frequency dependence of $\epsilon$ in our model.

\acknowledgments%%%%%%%%%%%%%%%%%%%%%%%%%%%%%%%%%%%%%%%% 

N.O.\ was supported by the Special Postdoctoral Researcher (SPDR) Program at RIKEN, the Incentive Research Project at RIKEN, 
and the Japan Society for the Promotion of Science (JSPS) Grants-in-Aid for Scientific Research (KAKENHI) Grants No.\ JP21K20371.
H.M.\ was supported by JSPS KAKENHI Grants No.\ JP18K13565 and No. \ JP22K03639. 

\appendix
\section{Resolution of our computation and the smallness of the cosmological constant}
\label{sec:app}
\begin{figure}[t]
  \begin{center}
    \includegraphics[keepaspectratio=true,height=75mm]{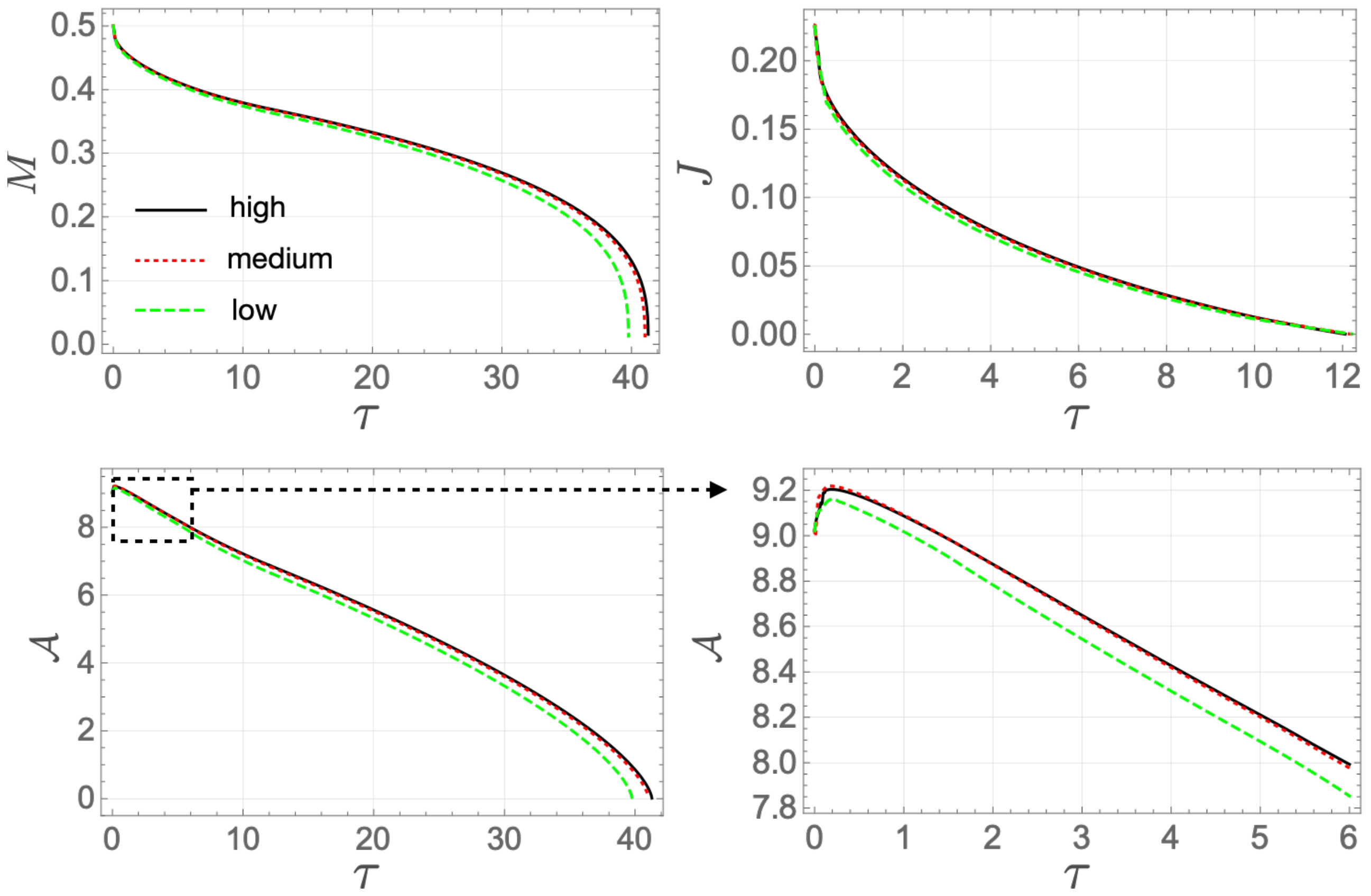}
  \end{center}
\caption{
Resolution test of our computation of $M(\tau)$, $J(\tau)$, and ${\cal A}(\tau)$. We take the resolution of $\{ N_i \} = (30,20,15,8,8)$ (high), $(20,15,10,5,5)$ (medium), and $(15,10,5,3,3)$ (low).
}
\label{resolutions}
\end{figure}

We performed the numerical computation to see the time development of the mass and angular momentum of a spinning black hole with the reflective boundary condition as shown in Figs. \ref{timeD} and \ref{all_timeD}. Here we perform a resolution test to check that our results shown in those figures were obtained with sufficiently high accuracy. We change the step size of the log-scaled spin parameter (introduced in Sec. \ref{sec:sub:time_evol}), $\Delta y$, as
\begin{align}
&\Delta y \equiv \frac{y_{i+1}-y_i}{N_{i}},\\
&\text{with} \ \{j_i\} = (j_{\rm max},0.7 j_{\rm max},0.4 j_{\rm max},0.1 j_{\rm max},0.01 j_{\rm max},0.001 j_{\rm max}),
\end{align}
where $y_i \equiv \ln j_i$, and we take $j_{\rm max} = 0.9$ throughout the paper. The results shown in Figs. \ref{timeD} and \ref{all_timeD} are obtained with $\{N_i \} = (20,15,10,5,5)$. Figure \ref{resolutions} shows our results obtained from high, medium, and low resolutions, and we find that the medium resolution we used in Figs. \ref{timeD} and \ref{all_timeD} is high enough.
%%%%%%%%%%%%%%%%%%%%%%%%%
%%%%%%%%%%%%%%%%%%%
%%%%%%%%%%%%%%%%%%%%%%%%%
\begin{figure}[t]
  \begin{center}
    \includegraphics[keepaspectratio=true,height=75mm]{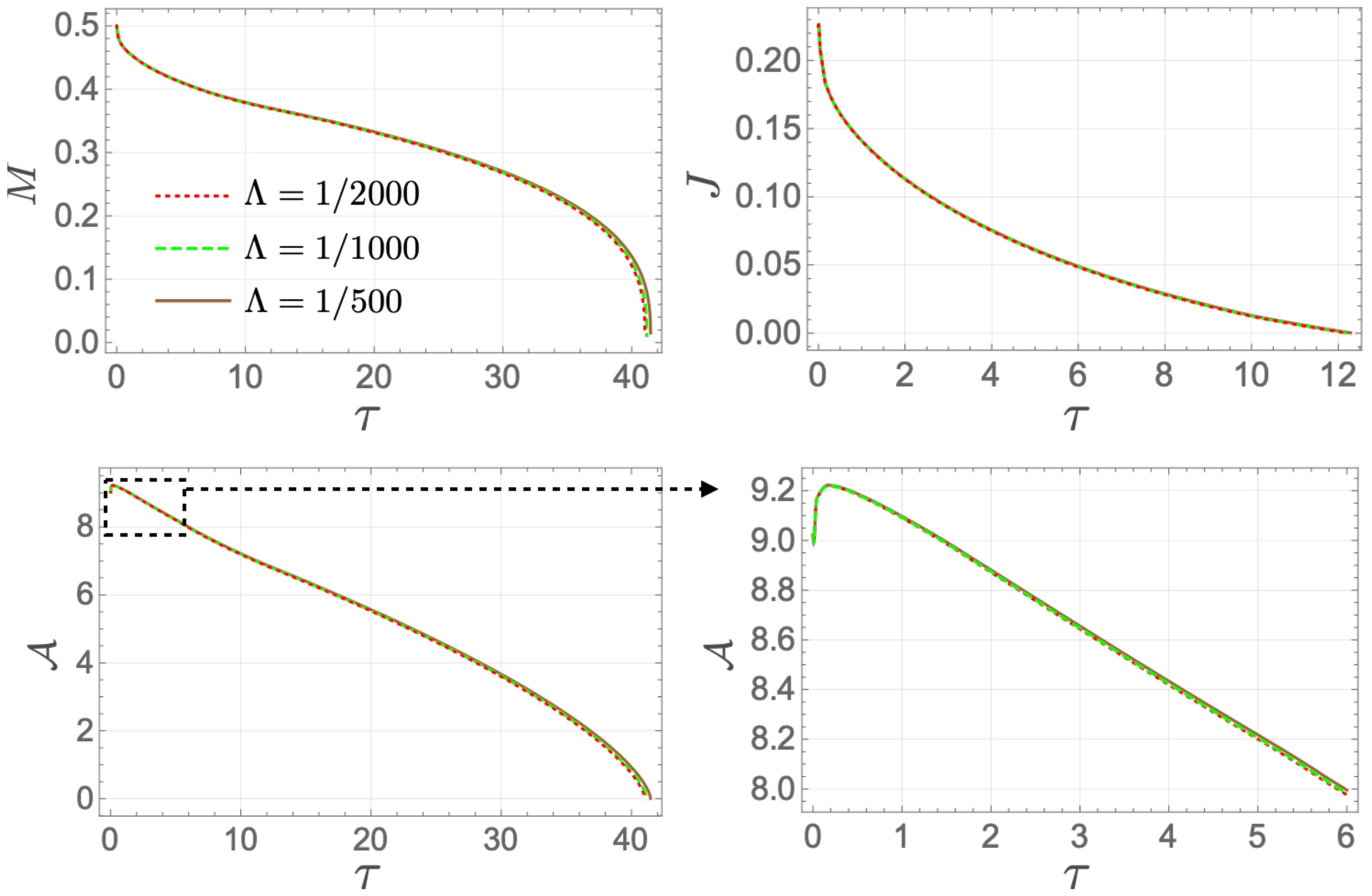}
  \end{center}
\caption{
Evolution of an echoing black hole [$M(\tau)$, $J(\tau)$, ${\cal A}(\tau)$] with a small cosmological constant of $\Lambda = 1/2000$, $1/1000$, and $1/500$.
}
\label{CC}
\end{figure}
%%%%%%%%%%%%%%%%%%%

We use the analytic solutions of the Heun's differential equation to obtain the graybody factor at the cost of having a small cosmological constant. Throughout the main text, we use $\Lambda = 1/2000$ with the normalization of the initial mass $M_o=1/2$ of the black hole. Figure \ref{CC} shows the time development of $M$, $J$, and ${\cal A}$ for various small cosmological constants: $\Lambda = 1/2000$, $1/1000$, $1/500$. We find that the result is insensitive to the small values of the cosmological constant at least for $\Lambda \leq 1/500$. Therefore, our result is valid for the Kerr black hole that is of our interest.
The deviation from the Kerr case may be significant for $\Lambda \gtrsim 0.1$ as the superradiance is strongly affected by the cosmological constant when $\Lambda \gtrsim 0.1$ with $M=M_o=1/2$ as was shown in Ref.~\cite{Gregory:2021ozs}.

\end{document}